\documentclass[11pt]{article}
\usepackage[margin=1in]{geometry}
\usepackage{graphicx}

\def\beq{\begin{equation}}
\def\eeq#1{\label{#1}\end{equation}}
\def\eeqn{\end{equation}}

\def\beqa{\begin{eqnarray}}
\def\eeqa#1{\label{#1}\end{eqnarray}}
\def\eeqan{\end{eqnarray}}

\let\bar=\overbar

\def\Dslash{\not{\hbox{\kern-4pt $D$}}}
\def\dslash{\not{\hbox{\kern-2pt $\del$}}}

\def\msb{{\bar{\ssstyle M \kern -1pt S}}}

\def\Title#1{\begin{center} {\Large {\bf #1} } \end{center}}
\def\Author#1{\begin{center} {\normalsize {\sc #1} } \end{center}}
\def\Institution#1{\begin{center} {\normalsize {\it #1} } \end{center}}
\def\Abstract#1{\noindent {\normalsize {\bf Abstract:} {\normalfont #1}}}
\def\Conference{\vspace{4mm}\begin{raggedright} {\normalsize {\it Talk presented at the 2019 Meeting of the Division of Particles and Fields of the American Physical Society (DPF2019), July 29--August 2, 2019, Northeastern University, Boston, C1907293.} } \end{raggedright}\vspace{4mm}}

\begin{document}

%
%

\Title{Low Noise Cold Electronics System for SBND LAr TPC}

\Author{Shanshan Gao, for the SBND collaboration}

\Institution{Physics Department\\ Brookhaven National Laboratory, Upton, NY 11973, USA}

\Abstract{The Short Baseline Near Detector (SBND) is one of three liquid argon (LAr) neutrino detectors sitting in the Booster Neutrino Beam (BNB) at Fermilab as part of the Short Baseline Neutrino (SBN) program. The detector is in a cryostat holding 260-ton of LAr and consists of four 2.5 m (L) $\times$ 4 m (W) Anode Plane Assembles (APAs) and two Cathode Plane Assemblies (CPAs), which leads to 11,264 Time Projection Chamber (TPC) readout channels and two separate 2 m long drift regions. As an enabling technology, Cold Electronics (CE) developed for cryogenic temperature operation makes possible an optimum balance among various design and performance requirements for such large sized detectors. Brookhaven National Laboratory (BNL) has been leading the R\&D and implementation of the entire front-end CE system for LAr TPC readout in collaboration with other SBND institutes. The front-end readout electronics system includes the cold front-end electronics placed close to the wire electrodes, which detects and digitizes the charge signal in LAr, as well as the warm interface electronics placed on the signal feed-through flange outside of the cryostat, which further organizes and transmits the digitized signal to the DAQ system. An extensive study of electronics suitable for 77 K - 300 K, including the custom designed front-end ASIC and commercial components, e.g. ADC and FPGA, has been made to meet requirements such as low noise, low power consumption, high reliability and long lifetime. Furthermore, an integral design concept of APA, CE, feed-through, warm interface electronics with local diagnostics, grounding and isolation rules has been practiced with vertical slice test stands to make projection of the CE performance in the SBND detector. }

\Conference

%
%

\section{Introduction}

\begin{figure}[!htb]
\centering
\includegraphics[width=4.5 in]{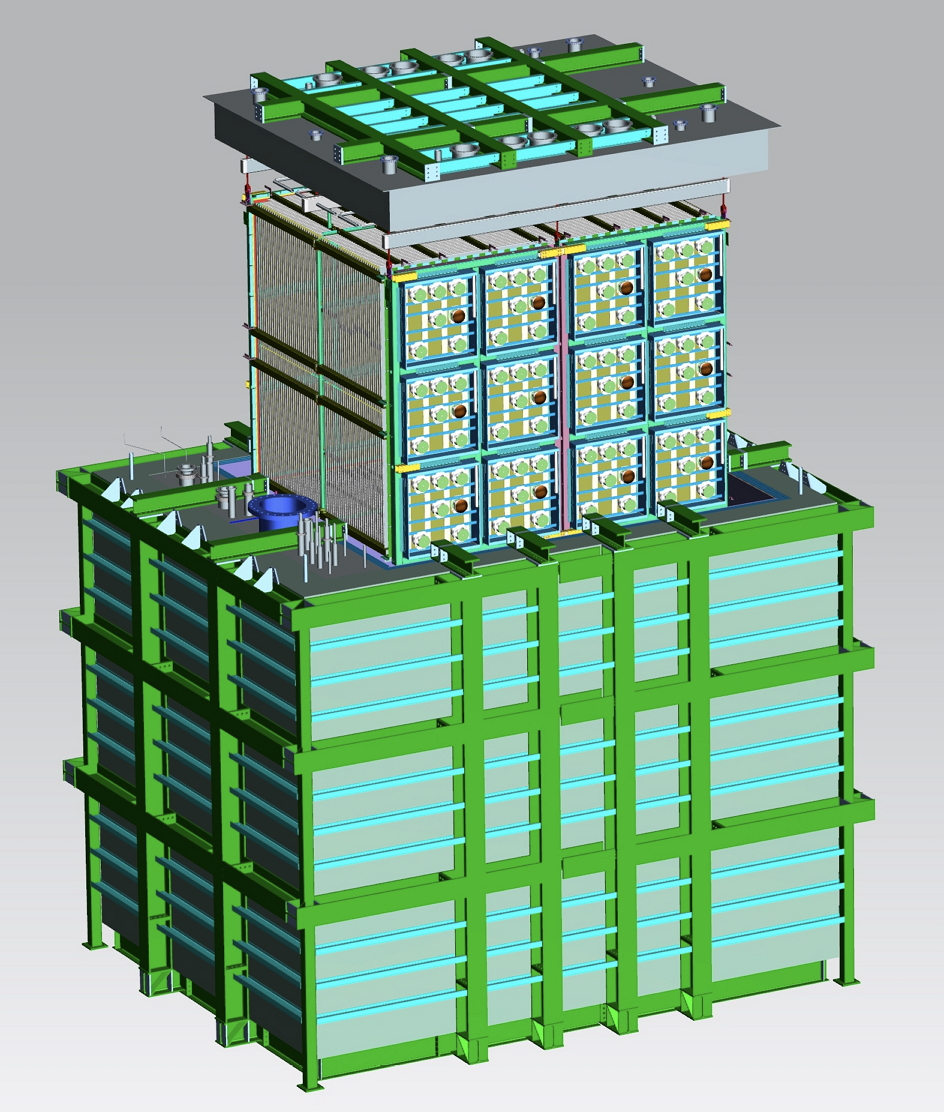}
\caption{Overview of SBND LAr TPC.  2 APAs are on either side; the sensing area of each APA is 2.5 $\times$ 4 m. With Two CPAs in central, two separate 2 m long drift regions are formed.}
\label{fig:fig1}
\end{figure}

\begin{figure}[!htb] 
\centering
\includegraphics[width=4.5 in]{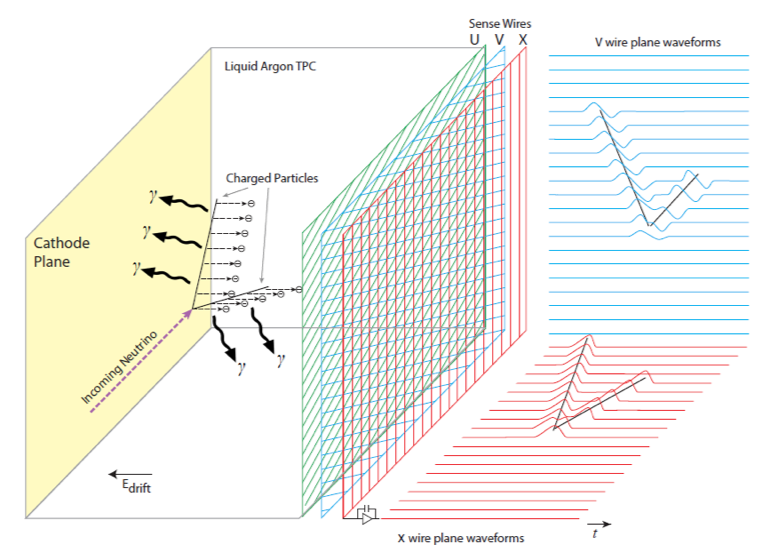}
\caption{The general operating principle of the single-phase LAr TPC.}
\label{fig:fig2}
\end{figure}

The Short Baseline Near Detector (SBND) is one of three liquid argon (LAr) neutrino detectors sitting in the Booster Neutrino Beam (BNB) at Fermilab as part of the Short Baseline Neutrino (SBN) program. The detector is in a cryostat holding 260-ton LAr and consists of four 2.5 m (L) $\times$ 4 m (W) Anode Plane Assembles (APAs) plus 2 Cathode Plane Assemblies (CPAs), as shown in Figure~\ref{fig:fig1}, which has $2 \times 2 m$ drift regions and 11, 264 TPC (Time Projection Chamber) readout channels \cite{paper1}. As shown in Figure~\ref{fig:fig2} is the general principle of the single-phase LAr TPC, a large volume of LAr is subjected to a strong electric field of a few hundred volts per centimeter. Charged particles passing through the detector ionize the argon atoms, and the ionization electrons drift in the electric field to the anode wall on a timescale of milliseconds. This anode consists of layers of active wires forming a grid. The relative voltage between the layers is chosen to ensure all but the final layer are transparent to the drifting electrons, and these first layers produce bipolar induction signals as the electrons pass through them. The final layer collects the drifting electrons, resulting in a unipolar signal \cite{paper2}. 

\begin{figure}[!htb]
\centering
\includegraphics[width=4.5 in]{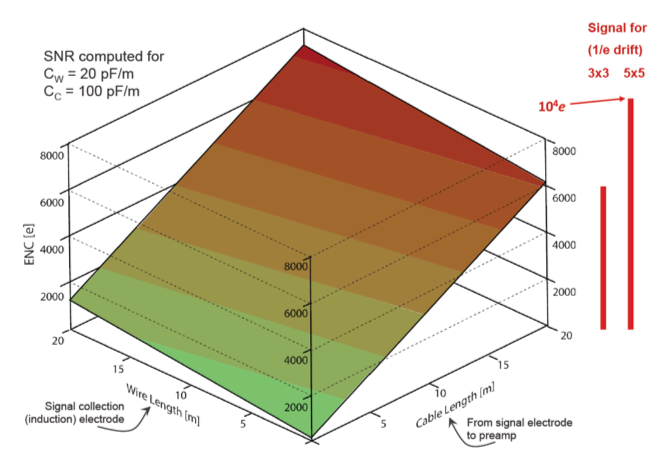}
\caption{Noise simulation with sensing wire length and cable length. Integrated front-end electronics close to detector electrodes yields the best SNR because the cable from signal electrodes to pre-amplifier is minimized.}
\label{fig:fig3}
\end{figure}

As an enabling technology, cold electronics developed for cryogenic temperature (77 K - 89 K) operation makes possible an optimum balance among various design and performance requirements for such large sized detectors \cite{paper3}. There are two main advantages of cold electronics. First, large detectors used for neutrino experiment requires very low noise to meet their physics requirements. Cold electronics decouples the electrode and cryostat design from the readout design. With electronics integral with detector electrodes, the input capacitance of signal cable is negligible, which results in the noise independent of the fiducial volume (signal cable lengths), as shown in Figure~\ref{fig:fig3}. Meanwhile, benefit from charge carrier mobility in silicon increasing and thermal fluctuations decreasing with kT/e, the noise of CMOS front end (FE) ASIC significantly decreases at cold temperature. Second, signal digitization and multiplexing to high speed links inside the cryostat result in large reduction in the quantity of cables (less outgassing) and the number of feed-through penetrations, also giving the designers of both the TPC and the cryostat the freedom to optimize the detector configurations. Therefore, SBND chooses cold electronics as LAr TPC readout solution. 

\section{Front-end Readout Electronics}

\begin{figure}[!htb]
\centering
\includegraphics[width=4.5 in]{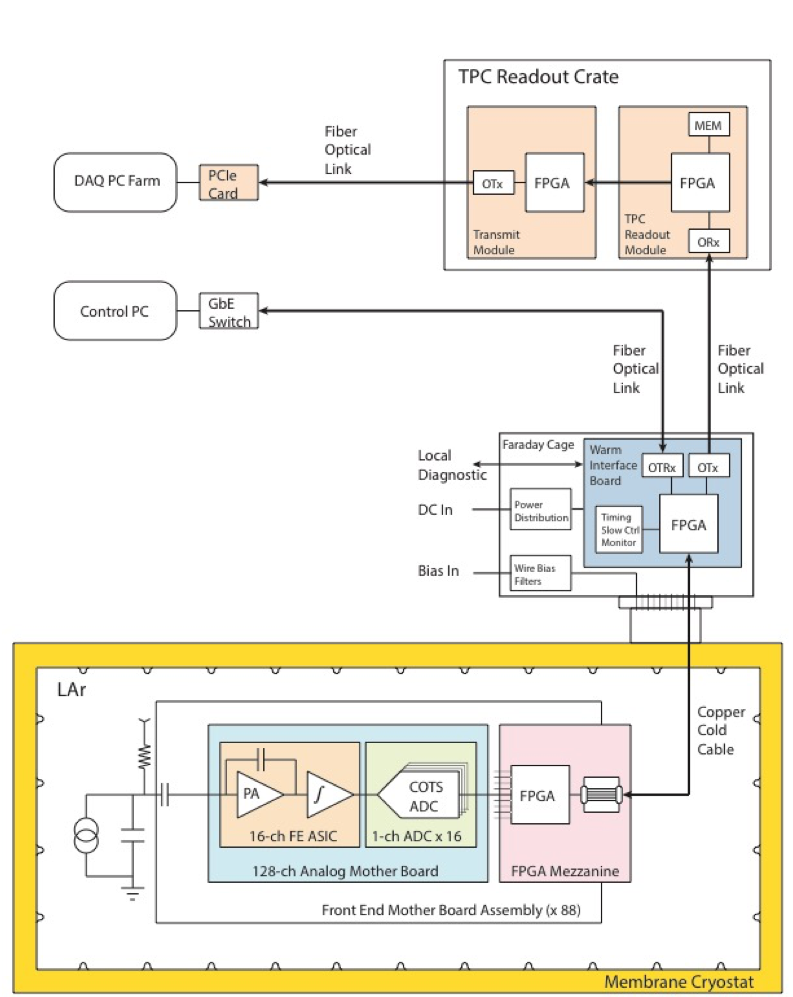}
\caption{Diagram of SBND TPC readout electronics. It consists of front-end electronics and back-end electronics. Front-end electronics, which is mainly developed and implemented by BNL, includes cold electronics located at detector electrodes inside the cryostat and warm interface electronics on the top of the signal feed-throughs outside the cryostat. Back-end electronics, also known as DAQ system, is implemented by Nevis Labs, Columbia University. }
\label{fig:fig4}
\end{figure}

Benefit from ProtoDUNE-SP front-end electronics development, the architecture of SBND front-end readout electronics, which is very similiar as ProtoDUNE-SP's, consists of cold electronics (CE) and warm interface electronics, as shown in Figure~\ref{fig:fig4} \cite{paper4}. Cold electronics, which is placed close to the detector electrodes inside the cryostat to get rid of signal cables and yield the best signal to noise ratio (SNR), consists of 88 Front End Mother Board (FEMB) assemblies to amplify, shape, digitize and multiplex 11, 264 TPC channels to warm interface electronics through 8 m cold copper cables. There are 4 sets of cold cable bundles connected to 4 sets of signal feed-through flanges. A Warm Interface Electronics Crate (WIEC) is installed on the top of each flange to hold warm interface electronics as a bridge between CE and DAQ, timing and slow control, power distribution systems. The DAQ, timing and slow control systems communicate with warm interface electronics through fiber optical links. This FE readout architecture is not only giving the freedom to the TPC and cryostat design with superior SNR, but also completely separating the FE from DAQ, timing and slow control systems to simplify the grounding and isolation design. Furthermore, with the local diagnostic function implemented in the warm interface electronics, the characterization of readout electronics, such as sense wire diagnostics, electronics calibration, and noise evaluation can be performed without DAQ and timing systems. This architecture makes it possible to have readout electronics development and testing independent of DAQ system. 

\subsection{Cold Electronics}
CMOS device at LAr temperature has much lower noise than at room temperature, however, there nearly isn’t any commercial semiconductor device designed for cryogenic temperature operation. We have to design the ASIC working at cryogenic temperature, or screen various commercial CMOS devices to find survivors at cryogenic temperature. 

BNL has started cryogenic CMOS ASIC R\&D since 2008. Studies conducted at BNL indicate charge carrier mobility in silicon increases and thermal fluctuations decrease with kT/e at 77 K - 89 K, resulting in a higher gain, higher $g_{m}/I_{D}$, higher speed and lower noise.  Channel Hot Carry Effect (HCE), which is the only remaining aging mechanism that affects the lifetime of CMOS devices at cryogenic temperature, is evaluated as well \cite{paper5, paper6, paper7}. A 16-channel FE ASIC which is suitable for 77 K - 300 K operation with long lifetime and low power consumption is designed. Each FE ASIC channel has a charge amplifier circuit with a programmable gain selectable from one of 4.7, 7.8, 14 and 25 mV/fC (full scale charge of 55, 100, 180 and 300 fC), a high-order anti-aliasing filter with programmable time constant (peaking time 0.5, 1, 2, and 3 $\mu$s), an option to enable AC coupling, and a baseline adjustment for operation with either the collection (200 mV) or the induction (900 mV) wires. In addition, each FE ASIC channel is equipped with an injection capacitor for calibration purpose which can be enabled or disabled. Shared among the 16 channels in the FE ASIC are the bias circuits, programming registers, a temperature monitor, an analog buffer for signal monitoring, and the digital interface. The power dissipation of FE ASIC is $\sim$6 mW/channel.

\begin{figure}[!htb]
\centering
\includegraphics[width=4.5 in]{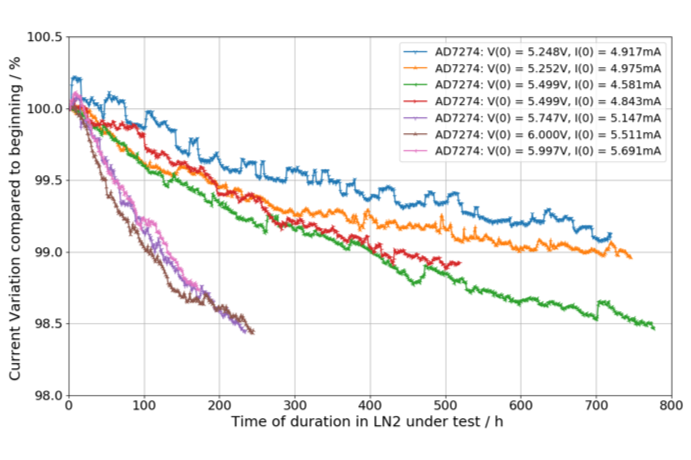}
\caption{Current of VDD vs. Time of duration in LN2. The current degradation is faster with higher stress voltages. }
\label{fig:fig5}
\end{figure}

\begin{figure}[!htb]
\centering
\includegraphics[width=4.5 in]{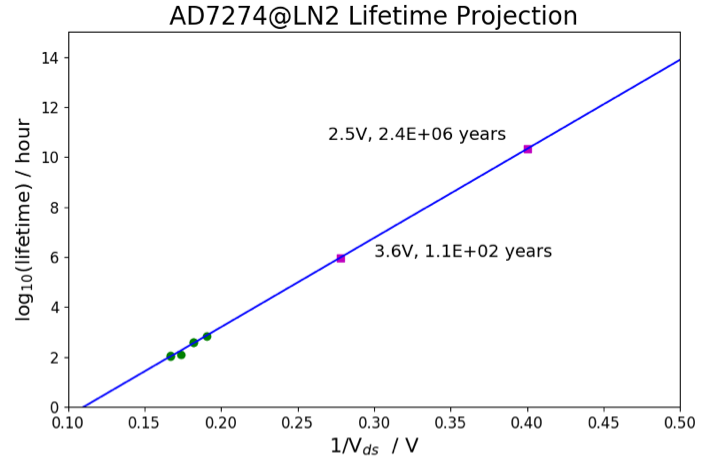}
\caption{A preliminary ADC lifetime projection. }
\label{fig:fig6}
\end{figure}

However, the cold ADC ASIC designed for ProtoDUNE-SP doesn’t meet the SBND requirement, we started to characterize the performance of Commercial-Off-The-Shelf (COTS) ADC chips at cryogenic temperature since June 2017. Several COTS ADC chips have been identified as good candidates for operation at cryogenic temperature after the initial screening test. One candidate, ADI AD7274 fabricated in TSMC 350 nm CMOS technology, of which performance at cold is same as at room temperature, DNL is 0.25 LSB and power consumption is less than 5 mW per ADC at 2 MS/s, is chosen to perform cryogenic lifetime measurement. The lifetime study of AD7274 at cryogenic temperature has been conducted by BNL and Manchester University since September 2017, which was concluded in about a year. As shown in Figure~\ref{fig:fig5}, current of VDD decreases consistently with different stress voltages. The current decreases faster with higher stress voltages, regardless of the test stand. Based on the past studies, lifetime due to HCE aging is a limit defined by a chosen level of monotonic degradation, so 1\% current drop is chosen as degradation criteria to project the life time though the ADC doesn’t fail at cold with the stress voltages. Based on the empirical equation: 
\begin{equation}
{log_{10}\tau \propto 1/V_{ds} }
\end{equation}
the preliminary ADC lifetime projection is plotted in Figure~\ref{fig:fig6}. The lifetime of AD7274 at the nominal 3.6V of 350 nm technology node is projected to $\sim$110 years. Reduced VDD to 2.5 V results in a very long extrapolated lifetime to  $\sim$\(2.4 \times 10^6\) years, which means HCE degradation is negligible. As a principle, the cold electronics for LAr TPCs should be designed for a lifetime one or more orders of magnitude longer than the required service life (e.g., $>100$ years for SBND). The lifetime study shows the AD7274 is cold qualified device to remain outside of the region of HCE degradation\cite{paper8}. As a milestone achievement, SBND collaboration made decision to use this ADC in 2018. 

Altera Cyclone IV FPGA and a low voltage (LV) regulator from TI passed the cryogenic qualification test done by BNL from various commercial CMOS device candidates. The FPGA has four 4:1 MUX circuits that combine the data from 128 ADC chips into 4 serial lines of 32 channels each, with four 1.28 Gbps serializers that drive the data in each line over cold cables to the warm interface electronics. The power dissipation of FPGA is about 1.75 W ($\sim$14 mW/channel). The LV regulator has ultra-low dropout voltage (typically 55 mV at 1.5 A) and low output noise ($<50$ $\mu$VRMS), making the device itself a good choice for cold electronics, which is very sensitive to the noise on power rail. 

\begin{figure}[!htb]
\centering
\includegraphics[width=4.5 in]{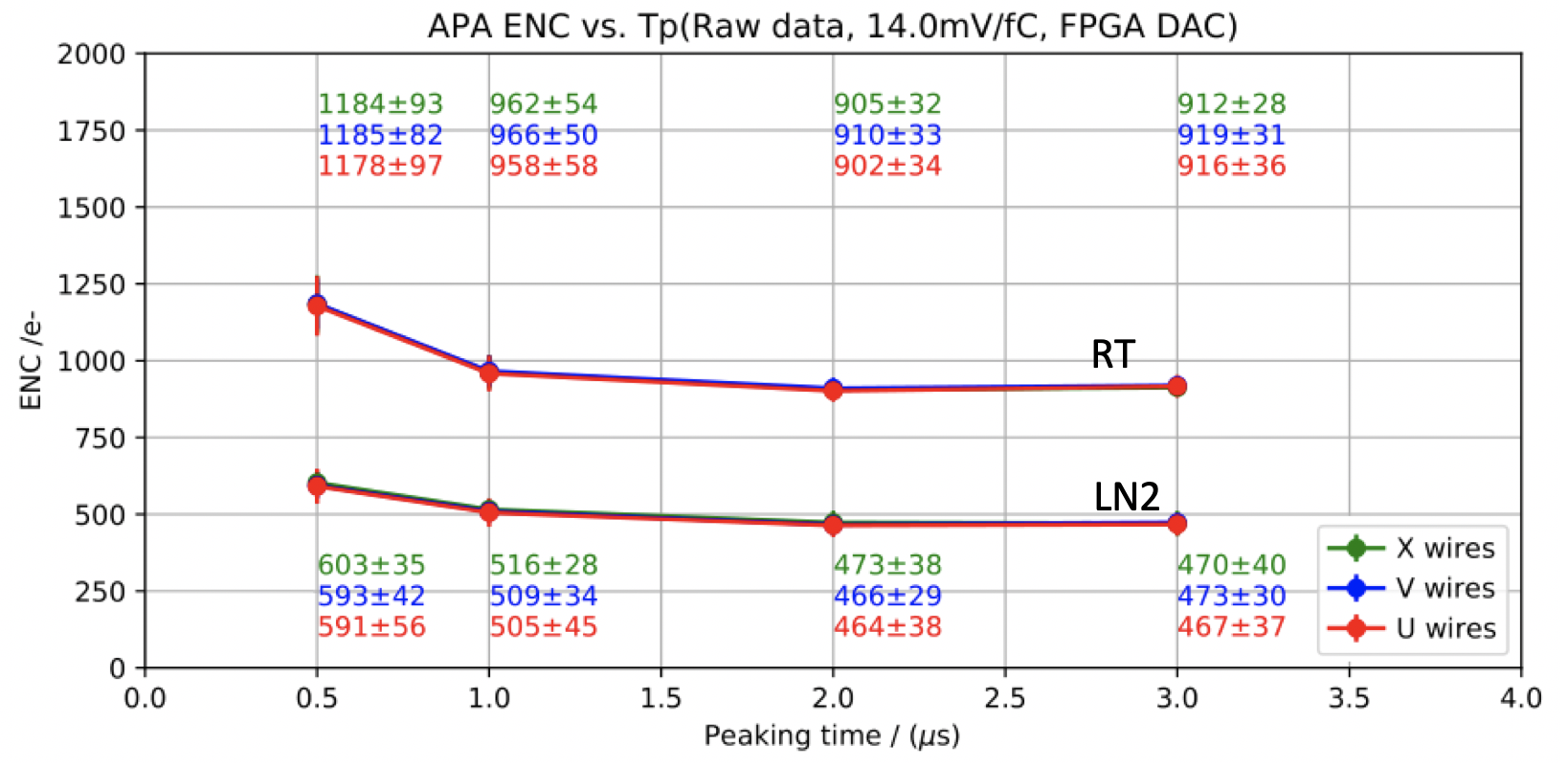}
\caption{Measured ENC as a function of filter-time constant (peaking time) with 150pF Cd. No matter the FE channels are configured for collection (X) wires or induction (V or U) wires, the noise only depends on the input capacitance.}
\label{fig:fig7}
\end{figure}

With aforementioned key components, a 128-channel FEMB has been designed as the key building block of the cold electronics attached to APA inside the cryostat. The total power consumption of each FEMB inside cryostat, including the power dissipation on 8 m cold power cable, is about 5.0 W. There is $\sim$10 mW/channel dissipated in power cables, cold regulators and RC filters.  With an on-chip test capacitor ($\sim$183 fF and stable at cryogenic temperature) for each channel, FE channel can be calibrated by on-chip 64 steps (6-bit DAC) pulse generator or external pulse sources. The nonlinearity of each channel is analyzed, which is less than 1\% as specified. To emulate the capacitance of APA sense wires (the capacitance of 7 m sense wire is approximately 150 pF in LAr), 150 pF MICA capacitor, of which capacitance has very small variation at warm and cold temperatures, is applied to the FE input. Noise characterization is made at both room and LN2 temperatures. Figure~\ref{fig:fig7} shows the measured ENC decreases significantly at cryogenic temperature. At LN2 temperature, the measured noise is around 470 $e^-$ at peaking time of 2.0 $\mu$s. 

\subsection{Warm Interface Electronics}
\begin{figure}[!htb]
\centering
\includegraphics[width=4.5 in]{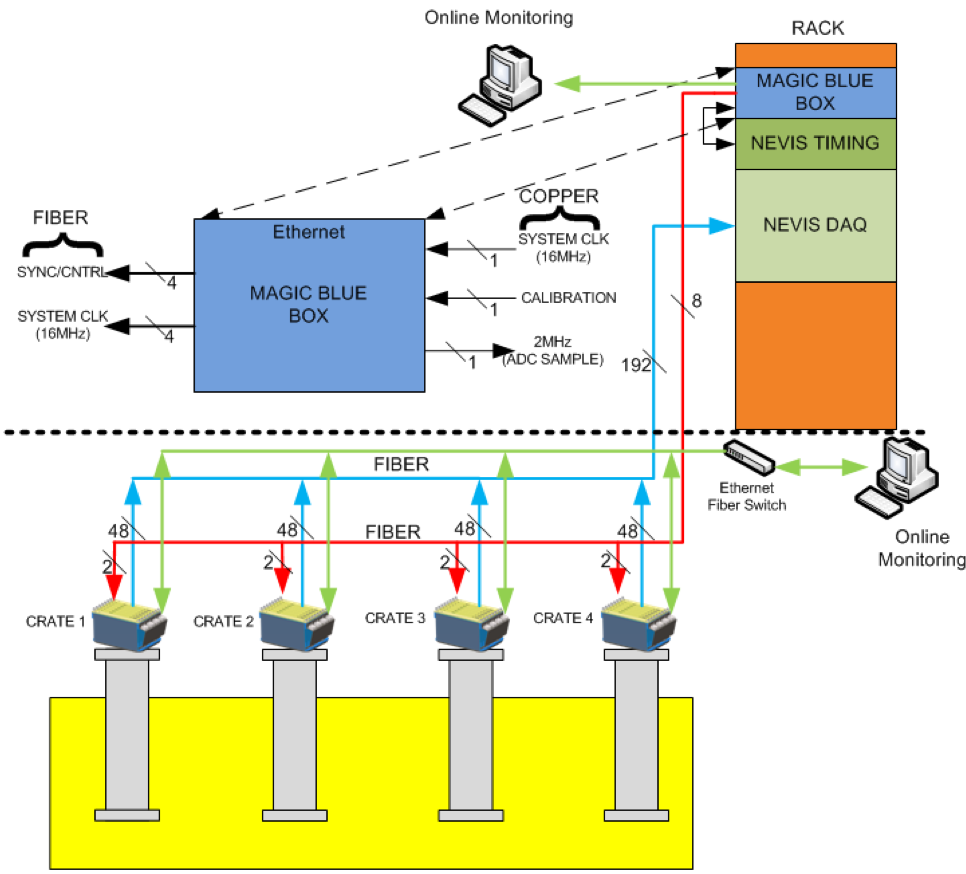}
\caption{Diagram of timing, control and data flow of warm interface electronics.}
\label{fig:fig8}
\end{figure}

Warm interface electronics, which is housed in WIEC attached directly to the signal flange, is the bridge between CE and DAQ system. Each WIEC is a faraday cage with only power and optical signals going in and out so that we can keep readout electronics from the unwanted pick-up noise. It includes one Power and Timing Card (PTC), up to six Warm Interface Boards (WIBs) and a passive Power and Timing Backplane (PTB). As shown in Figure~\ref{fig:fig8} , the PTC provides a bidirectional fiber interface to the timing system, and fans out clock and control signals to the six WIBs through PTB. A WIB receives the system clock and control signals and fan out those signals to four FEMBs. Meanwhile, it receives the high-speed data signals from the four FEMBs and transmits them to the DAQ system over fiber optical links. Warm interface electronics takes charge of power distribution and monitoring as well. The PTC receives 48 V power and steps down to 12 V, through PTB to feed WIBs. The WIB generates independent LV power for each FEMB and distributes them over cold cables. In addition, a Magic Blue Box (MBB) installed in the rack of Nevis back-end electronics distributes control and synchronized timing signals to each WIEC.

\section{Integration test}
SBND TPC uses extremely sensitive electronics to measure the small charge signals from the sense wires. As a necessary but not sufficient condition to achieve a good noise performance, the integral design concept of APA, CE, feed-through and warm interface electronics with local diagnostics, and strict grounding and isolation rules must be followed. Following the experience from ATLAS and MicorBooNE experiments, a grounding scheme is developed. The common of FE ASIC chips and of the rest components are connected to the common of FEMB which is attached to the APA frame over adapter boards. Through cold data and power cables, the common of FEMBs is connected together at the signal feed-through flange. In other words, all the electrical connections (power and signal) from APA and CE inside the cryostat only lead to the signal feed-through with 8 m cold cables. As a result, the feed-through is the only connection of the APA frame to the cryostat. At the warm side of feed-through, each WIEC is fed by a 48 V floating power supply. The cryostat ground is separated from the safety ground by a transformer to assure clean and low impedance grounding. To avoid potential ground loops, data transmission between warm interface electronics and other subsystems (DAQ, timing, and slow control) is only through fiber optical links\cite{paper9}. 

\begin{figure}[!htb]
\centering
\includegraphics[width=4.5 in]{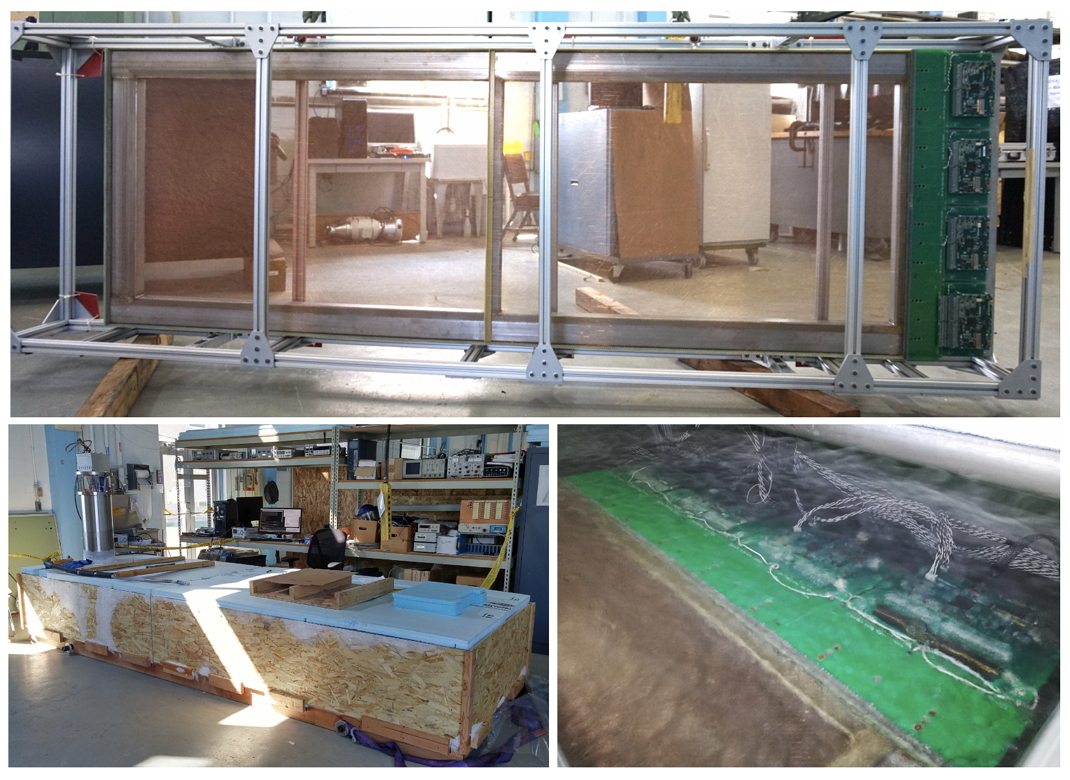}
\caption{The integration of APA and front end readout electronics. Top: 40\% APA instrumented with FEMB assemblies. Bottom left: Integration cold test with a cold box housing the 40\% APA. Bottom right: 40\% APA and cold electronics fully submerged in LN2. }
\label{fig:fig9}
\end{figure}

\begin{figure}[!htb]
\centering
\includegraphics[width=4.5 in]{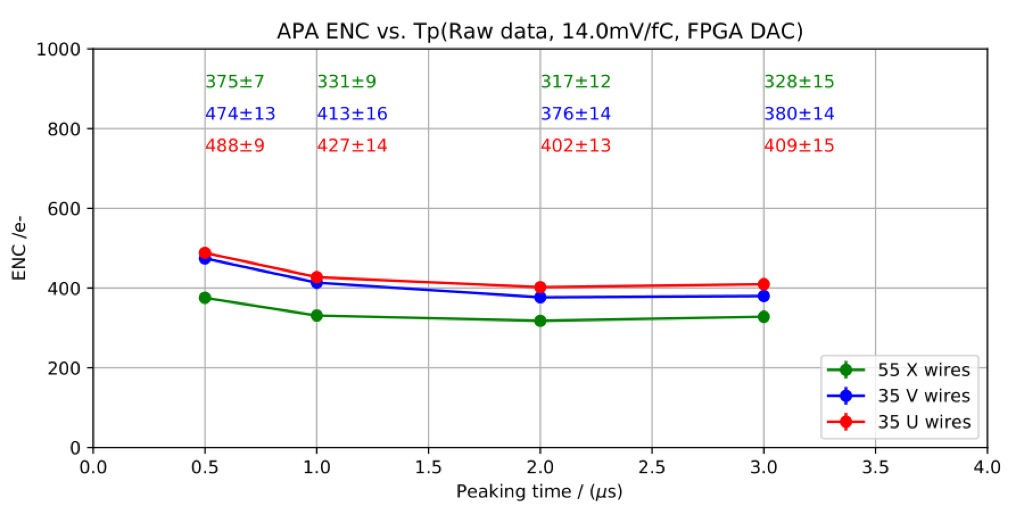}
\caption{40\% APA test result at LN2. The ENC of induction plane (U or V) is $\sim$400 $e^-$ and of collection plane (X) is $\sim$320 $e^-$ at 2 $\mu$s peaking time. }
\label{fig:fig10}
\end{figure}

\begin{figure}[!htb]
\centering
\includegraphics[width=4.5 in]{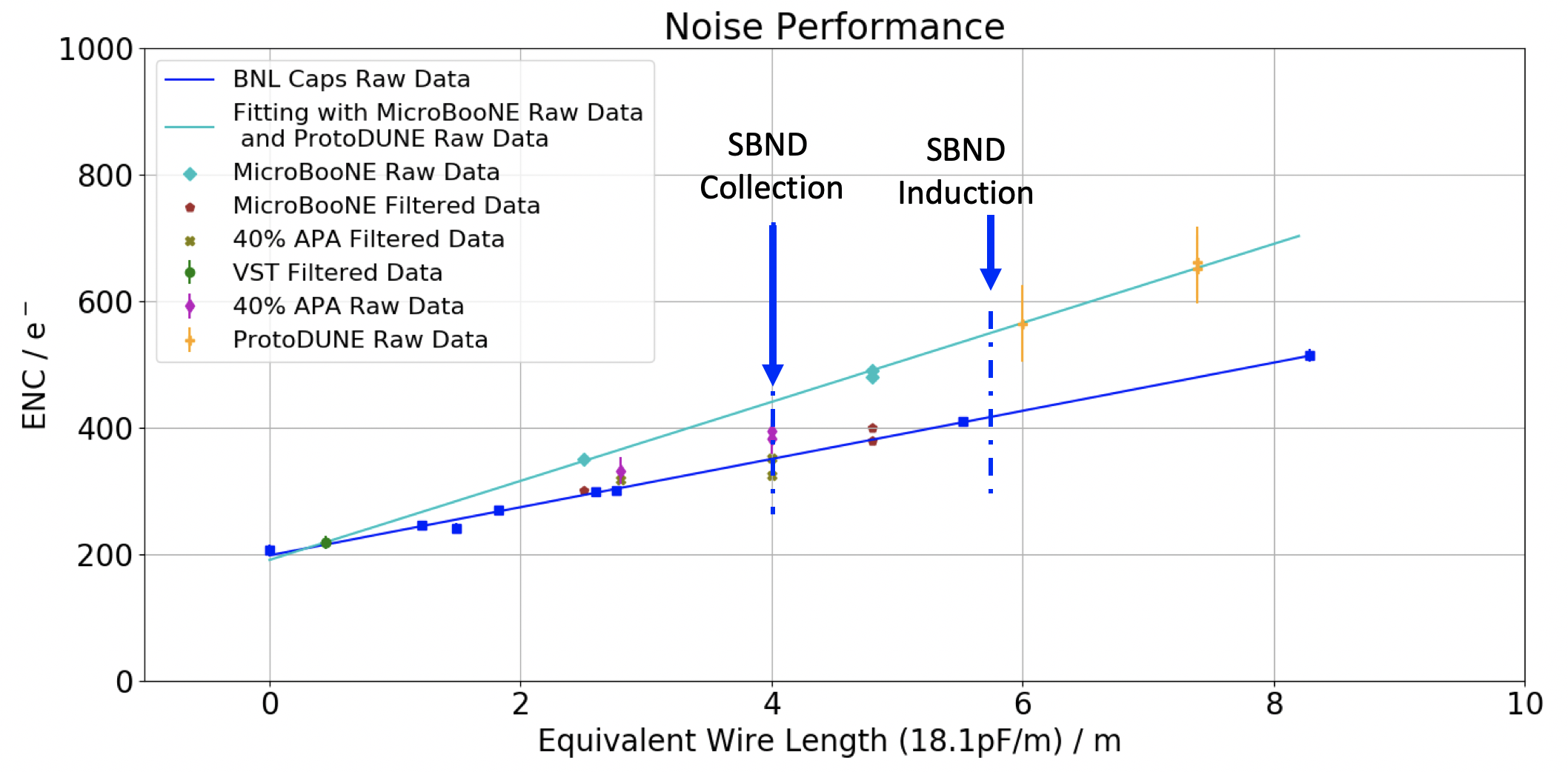}
\caption{ENC projection for SBND. Expected SBND ENC before offline filtering, induction plane is less 600 $e^-$, collection plane is less than 500 $e^-$. Offline noise filtering can get rid of coherent noise effectively, the best ENC can be achieved is only depending on the input capacitance. 
 }
\label{fig:fig11}
\end{figure}

A 40\% APA integration test stand is built at BNL. As shown in Figure~\ref{fig:fig9}, the APA sense area is about 2.8 $m^2$ with 2.8 m X-plane collection wires and 4.0 m U- and V-plane induction wires. It has 1,024 sense wires, can house up to 8 FEMB assemblies\cite{paper10}. Following the grounding rules mentioned above, 40\% APA instrumented with 4 SBND FEMB assemblies is submerged in liquid nitrogen completely. FEMBs are powered, configured, and read out by the warm interface electronics located in the WIEC on the top of signal feed-through chinmey through 8 m cold cables. The noise performance of 40\% APA submerged in LN2 is quite promising. The ENC of induction plane (U or V) is $\sim$400 $e^-$ and of collection plane (X) is $\sim$320 $e^-$ at 2 $\mu$s peaking time, as shown in Figure~\ref{fig:fig10}. Based on the test results and data collected from MicroBooNE and ProtoDUNE-SP experiments, a noise projection is made to predict the noise performance of SBND as shown in Figure~\ref{fig:fig11} . The noise of the induction plane is expected to be $<600$ $e^-$ and collection plane is $<500$ $e^-$ before applying offline filtering.

\section{Conclusion}
Readout electronics developed for low temperatures (77 K - 89 K) is an enabling technology for noble liquid detectors for neutrino experiments due to much lower noise and less cryostat penetrations. Benefit from ProtoDUNE-SP CE development, SBND cold electronics development is progressing well. A COTS ADC with negligible degradation caused by HCE at cryogenic temperature is identified and applied in SBND cold electronics with expected performance. 
The integration test of 40\% APA at BNL shows satisfactory noise performance, as well as ENC projection shows cold electronics meets SBND requirement. Currently, SBND front-end electronics production is completed as planed. With continuous effort for installation in the coming months and we are looking forward to commissioning in 2020.

\end{document}